\documentclass[epj]{svjour}
\usepackage{graphicx}
\begin{document}
\title{Photoproduction of $\omega$ mesons on nuclei near the production threshold}
\author{M.~Nanova$^{1}$, J.~Weil$^{2}$, S.~Friedrich$^{1}$, V.~Metag$^{1}$, U.~Mosel$^{2}$, M.~Thiel$^{1}$, G.~Anton$^{3}$, J.C.S.~Bacelar$^{4}$, O.~Bartholomy$^5$, D.~Bayadilov$^{5,6}$, Y.A.~Beloglazov$^6$, R.~Bogend\"orfer$^3$,~R.~Castelijns$^{4}$, V.~Crede$^{5,a}$, H.~Dutz$^7$, A.~Ehmanns$^5$, D.~Elsner$^7$, K.~Essig$^5$, R.~Ewald$^7$,~I.~Fabry$^5$, M.~Fuchs$^5$, Ch.~Funke$^5$, R.~Gothe$^{7,b}$, R.~Gregor$^1$, A.B.~Gridnev$^6$, E.~Gutz$^5$, S.~H\"offgen$^5$,~P.~Hoffmeister$^5$, I.~Horn$^5$, J. H\"ossl$^3$, I.~Jaegle$^8$, J.~Junkersfeld$^5$, H.~Kalinowsky$^5$,~Frank~Klein$^7$,~Friedrich~Klein$^7$, E.~Klempt$^5$, M.~Konrad$^7$, B.~Kopf$^{9,10}$, M.~Kotulla$^1$, B.~Krusche$^8$, J.~Langheinrich$^{7,10}$, H.~L\"ohner$^4$, I.V.~Lopatin$^6$, J.~Lotz$^5$, S.~Lugert$^1$, D.~Menze$^7$, T.~Mertens$^8$, J.G. Messchendorp$^4$, C.~Morales$^7$,~R.~Novotny$^1$, M.~Ostrick$^{7,c}$, L.M.~Pant$^{1,d}$, H.~van Pee$^5$, M.~Pfeiffer$^1$, A. Roy$^{1,e}$, A.~Radkov$^6$,~S.~Schadmand$^{1,f}$, Ch.~Schmidt$^5$, H.~Schmieden$^7$, B.~Schoch$^7$, S.~Shende$^{4,h}$, G. Suft$^3$,~A.~S\"ule$^7$, V.~V.~Sumachev$^6$, T.~Szczepanek~$^5$, U.~Thoma$^5$, D.~Trnka$^1$, R.~Varma$^{1,g}$, D.~Walther$^5$, Ch.~Weinheimer$^{5,i}$,  Ch.~Wendel$^5$\\
 (The CBELSA/TAPS Collaboration)
 \mail{{Mariana.Nanova@physik.uni-giessen.de}}
 }
\titlerunning{Photoproduction of $\omega$ mesons on nuclei near the production threshold}
\authorrunning{~M.~Nanova et al.}
\institute{%
$^{1}$II. Physikalisches Institut, Universit\"at Gie{\ss}en, Germany\\
$^{2}$Inst. f\"ur Theoretische Physik I, Universit\"at Gie{\ss}en, Germany\\
$^{3}$Physikalisches Institut, Universit\"at Erlangen, Germany\\
$^{4}$Kernfysisch Versneller Institut  Groningen, The Netherlands\\
$^{5}$Helmholtz-Institut f\"ur Strahlen- u. Kernphysik Universit\"at Bonn, Germany \\
$^6$Petersburg Nuclear Physics Institute Gatchina, Russia\\
$^{7}$Physikalisches Institut, Universit\"at Bonn, Germany\\
$^{8}$Physikalisches Institut, Universit\"at Basel, Switzerland\\
$^{9}$Institut f\"ur Kern- und Teilchenphysik TU Dresden, Germany\\
$^{10}$Physikalisches Institut, Universit\"at Bochum, Germany\\
$^a$Current address: Florida State University Tallahassee, FL, USA\\
$^b$Current address: University of South Carolina Columbia, SC, USA\\
$^c$Current address: Physikalisches Institut Universit\"at Mainz, Germany\\
$^d$Current address: Nuclear Physics Division BARC, Mumbai, India\\
$^e$Current address: IIT Indore, Indore, India\\
$^f$Current address: Forschunszentrum J\"ulich, Germany\\
$^g$Current address: Department of Physics I.I.T. Powai, Mumbai, India\\
$^h$Current address: Department of Physics, University of Pune, Pune, India\\
$^i$Present address: Universit\"at M\"unster, Germany\\
}%
\date{Received: date / Revised version: date}
%
\abstract{
The photoproduction of $\omega$ mesons on $LH_2$, $C$ and $Nb$ has been measured for incident photon energies from 900 to 1300 MeV using the CB/TAPS detector at ELSA. The 
$\omega$ line shape does not show any significant difference between the $LH_2$ and the $Nb$ targets. The experiment was motivated by transport calculations that predicted a sensitivity of the $\omega$ line shape to in-medium modifications near the production threshold on a free nucleon of $E_{\gamma}^{lab}$ = 1109 MeV. A comparison with recent calculations is given.}


\maketitle

\section{Introduction}
\label{intro}
Modifications of hadron properties in a strongly interacting environment have attracted a
lot of attention and have been intensively studied both theoretically and
experimentally. These studies were motivated by the expectation that chiral
symmetry, a fundamental symmetry of Quantum Chromodynamics (QCD) in the limit of vanishing
quark masses, may be restored in a nuclear medium at high temperatures or
\begin{figure*}
\resizebox{1.0\textwidth}{!}
 {
  \includegraphics{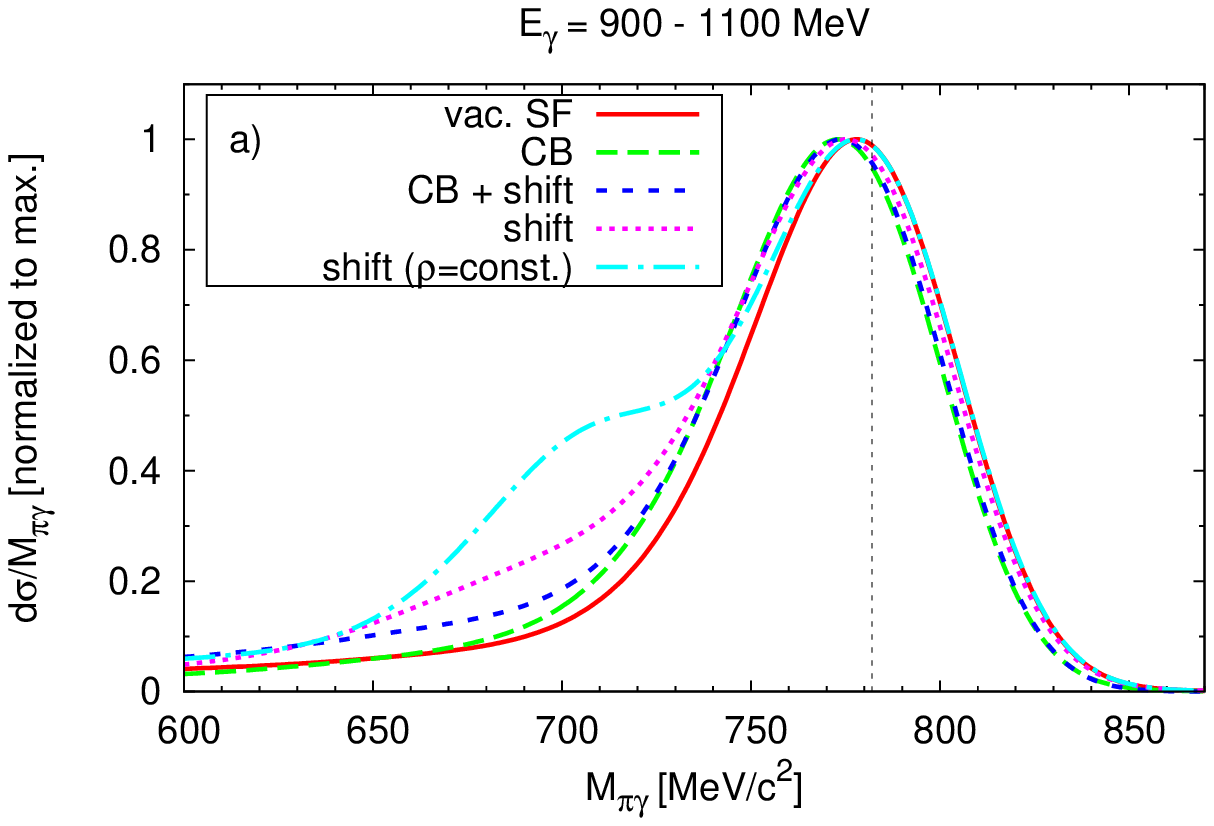}
  \includegraphics{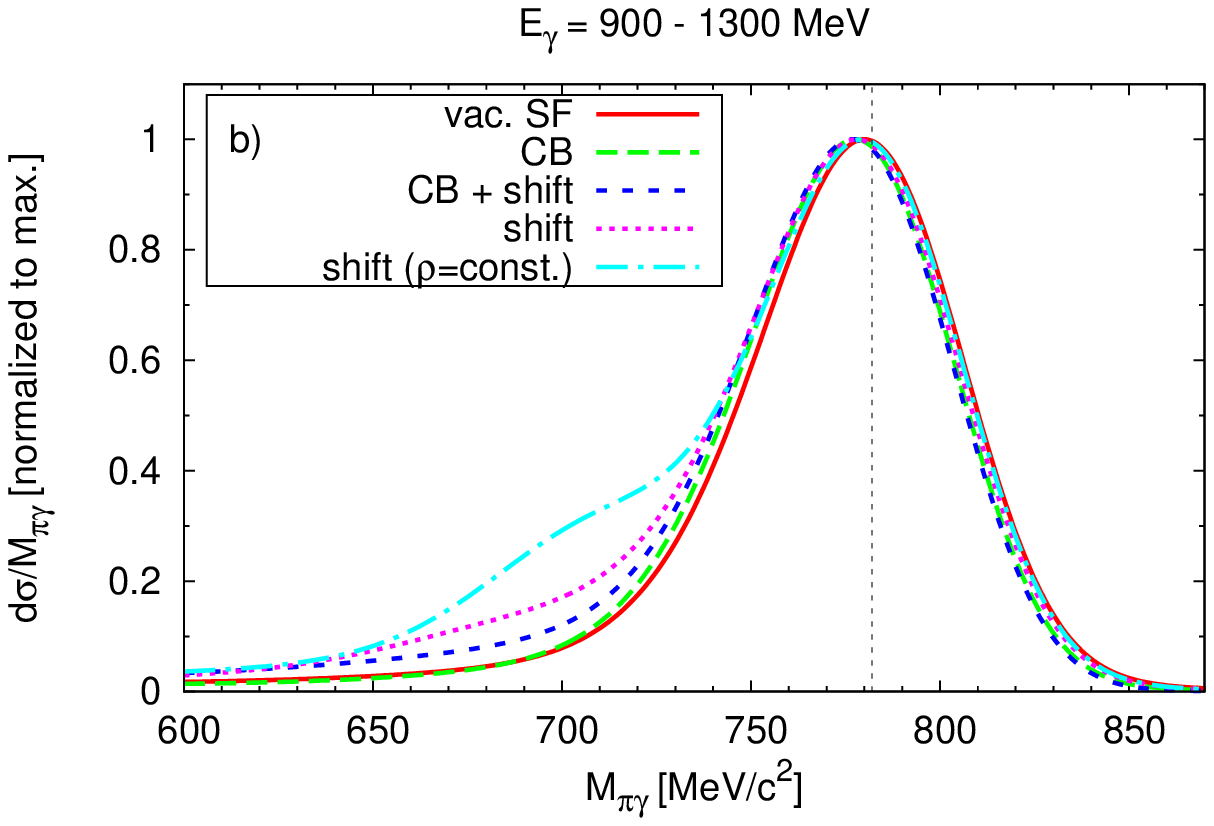}
 }
\caption{(Color online) $\omega$ meson line shape predicted for a $Nb$ target in GiBUU
 transport model calculations for different in-medium modification scenarios:
 vacuum spectral function (solid), collisional broadening of $\Gamma_{coll}$ = 140 MeV (long dashed),
 collisional broadening and an attractive mass shift of -16$\%$ at nuclear matter
 density (short dashed) and mass shift without broadening (dotted). The dashed-dotted
 curve shows the results for a constant nuclear density of $\rho = 0.6 \rho_0$. 
 The signals are folded with the detector response given by 
 Eq. \ref{novis_1} with the parameters $\sigma = $ 25 MeV and $\tau$ = - 0.09 and take into account a $1/E_{\gamma}$ weighting of the bremsstrahlung spectrum. 
 a) incident photon energies of 900 - 1100 MeV; b) incident photon energies of 900 - 1300 MeV.}
\label{fig:GiBUU}
\end{figure*}
densities. In vacuum this symmetry is broken as visible in the low mass part
of the hadronic spectrum: chiral partners - hadronic states with the same spin
but opposite parity - like the $\rho$ and $a_1$ meson are different in
mass while they should be mass degenerate if chiral symmetry were to hold.
It turns out, however, (for details see \cite{LMM}) that a connection between
chiral symmetry restoration and hadronic in-medium spectral functions is
much more involved. QCD sum rules provide a link between the quark-gluon
sector and hadronic descriptions but do not fix the properties of hadrons
in the strongly interacting medium. They only provide constraints for hadronic models
which are still needed for calculating the in-medium self-energies of hadrons
and their spectral functions.

Experimentally this field has been addressed in heavy ion collisions and
reactions with proton and photon beams. Light vector mesons
are particularly suited for these investigations since - after production in
a nuclear reaction - they decay in the nuclear medium with sizable probability
because of their short lifetimes. Experimental results are summarized and
critically evaluated in recent reviews \cite{LMM,Hayano_Hatsuda}.
Almost all experiments report a softening of
the spectral functions of the light vector mesons $\rho, \omega$ and
$\phi$. Increases in width are observed depending on the density and
temperature of the hadronic environment. Mass shifts are only reported by the
KEK group \cite{Naruki,Muto} who studied $\rho, \omega$ and $\phi$ production
in proton nucleus reactions at 12 GeV. The claim of a mass shift of the $\omega$
meson in photoproduction on $Nb$ \cite{Trnka} has not been confirmed in a
re-analysis of the data \cite{Nanova}.

In the latter experiment incident photon energies covered the range from 900 -
2200 MeV. Because of the increase of the production cross section with
photon energy most of the observed $\omega$ mesons are produced with photons
of energy larger than 1500 MeV. For the energy range of 1500 - 2200 MeV
transport calculations \cite{Gallmeister,Muehlich_PhD} have shown that the
$\omega$ lineshape is rather insensitive to different in-medium modification scenarios like
broadening or broadening and mass shift
since most of the $\omega$ decays occur
outside of the nuclear medium, even despite of a cut on the $\omega$
momentum ($p_{\omega} \leq 500$ MeV/c). Furthermore, due to the
experimentally observed strong absorption of $\omega$ mesons in the nuclear
medium \cite{Kotulla} $\omega$ mesons produced in the interior of the nucleus
are largely removed by inelastic reactions and do not reach the detector;
information on possible in-medium mass shifts thereby gets lost. The limited
sensitivity of the {$\omega$} lineshape to in-medium modifications has been
confirmed experimentally in \cite{Nanova}.

\label{sec:exp}
\begin{figure*}
 \resizebox{0.9\textwidth}{!}{
   \includegraphics{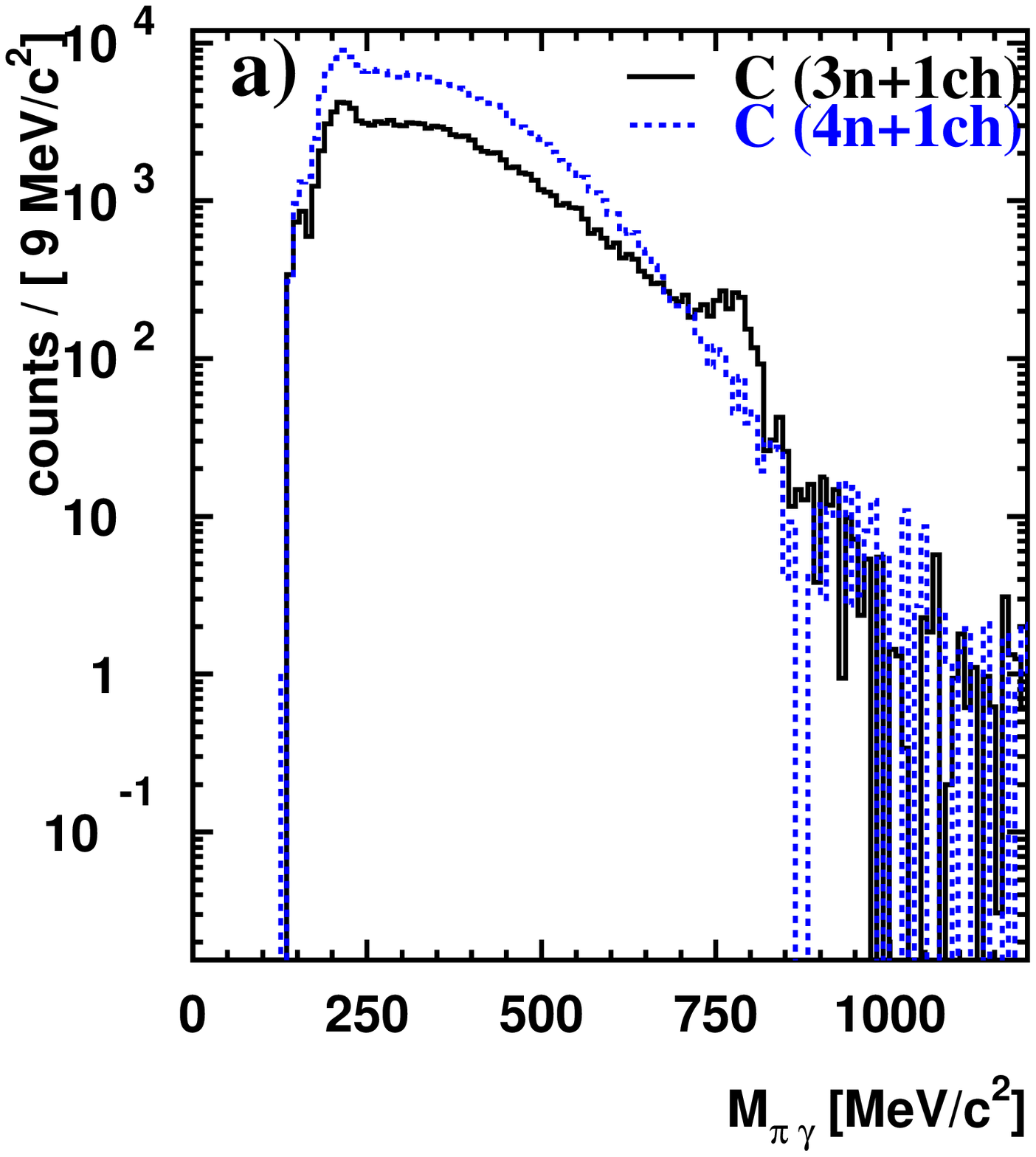} \includegraphics{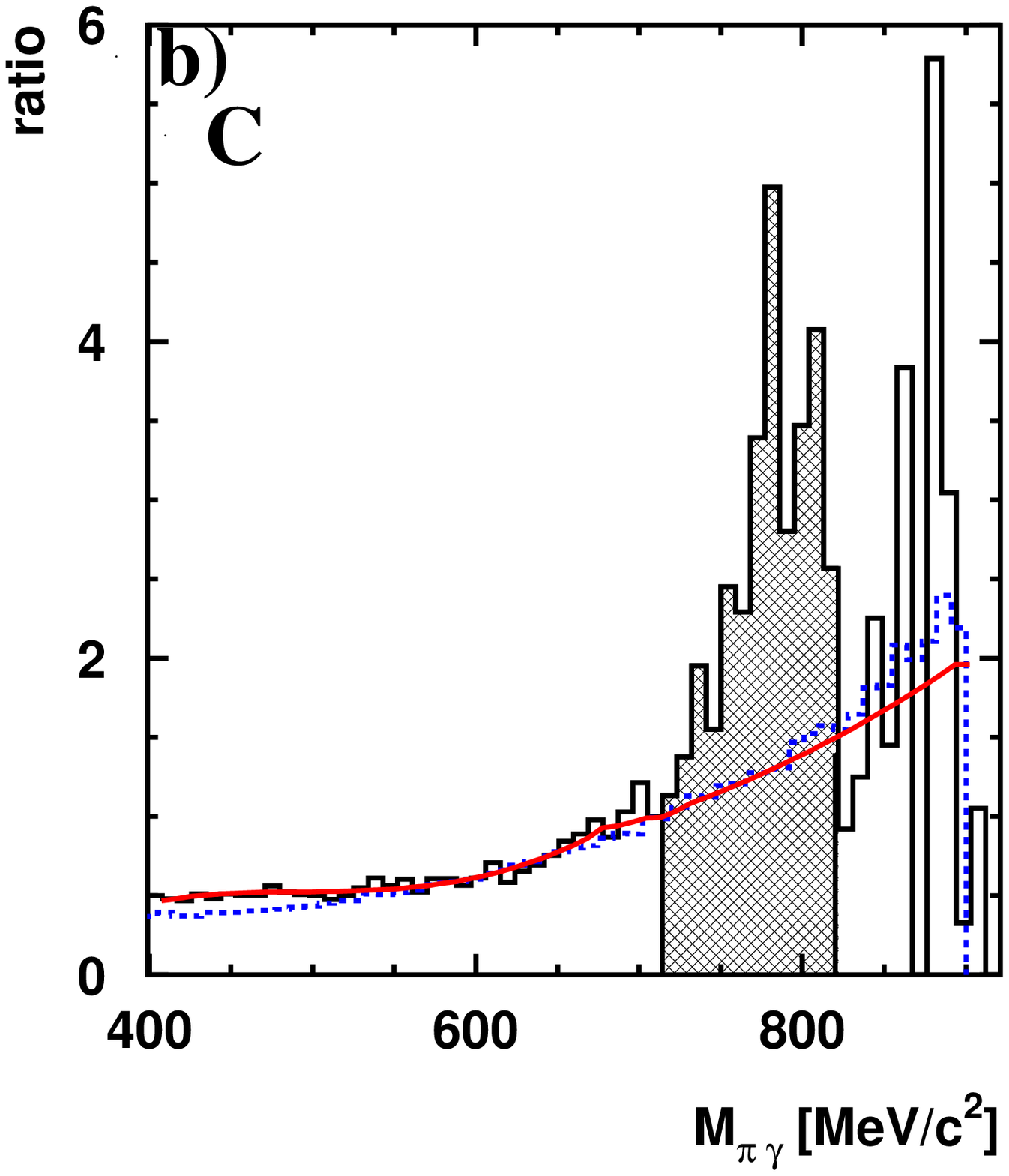}}
 \resizebox{0.9\textwidth}{!}{
    \includegraphics{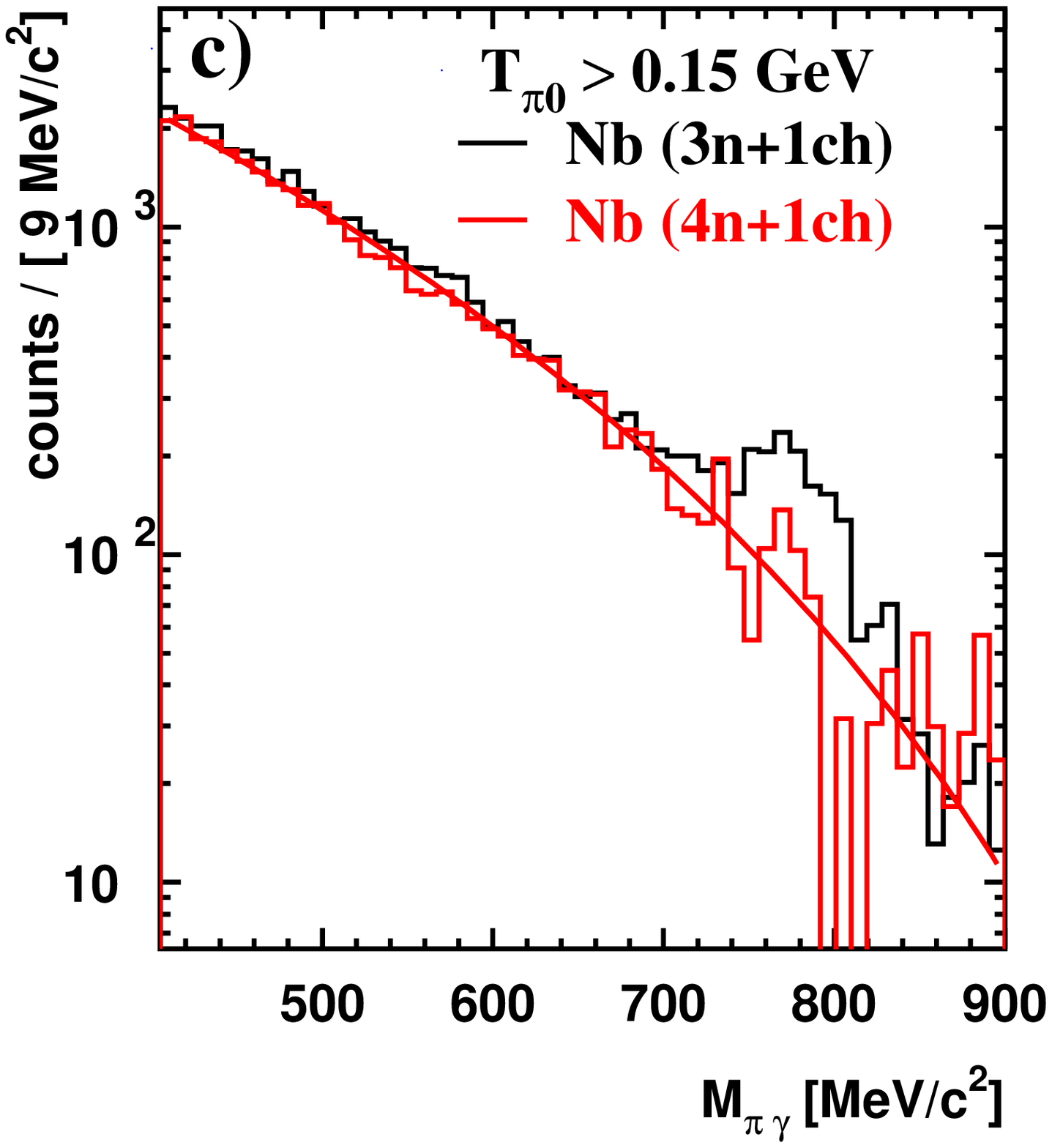}\includegraphics{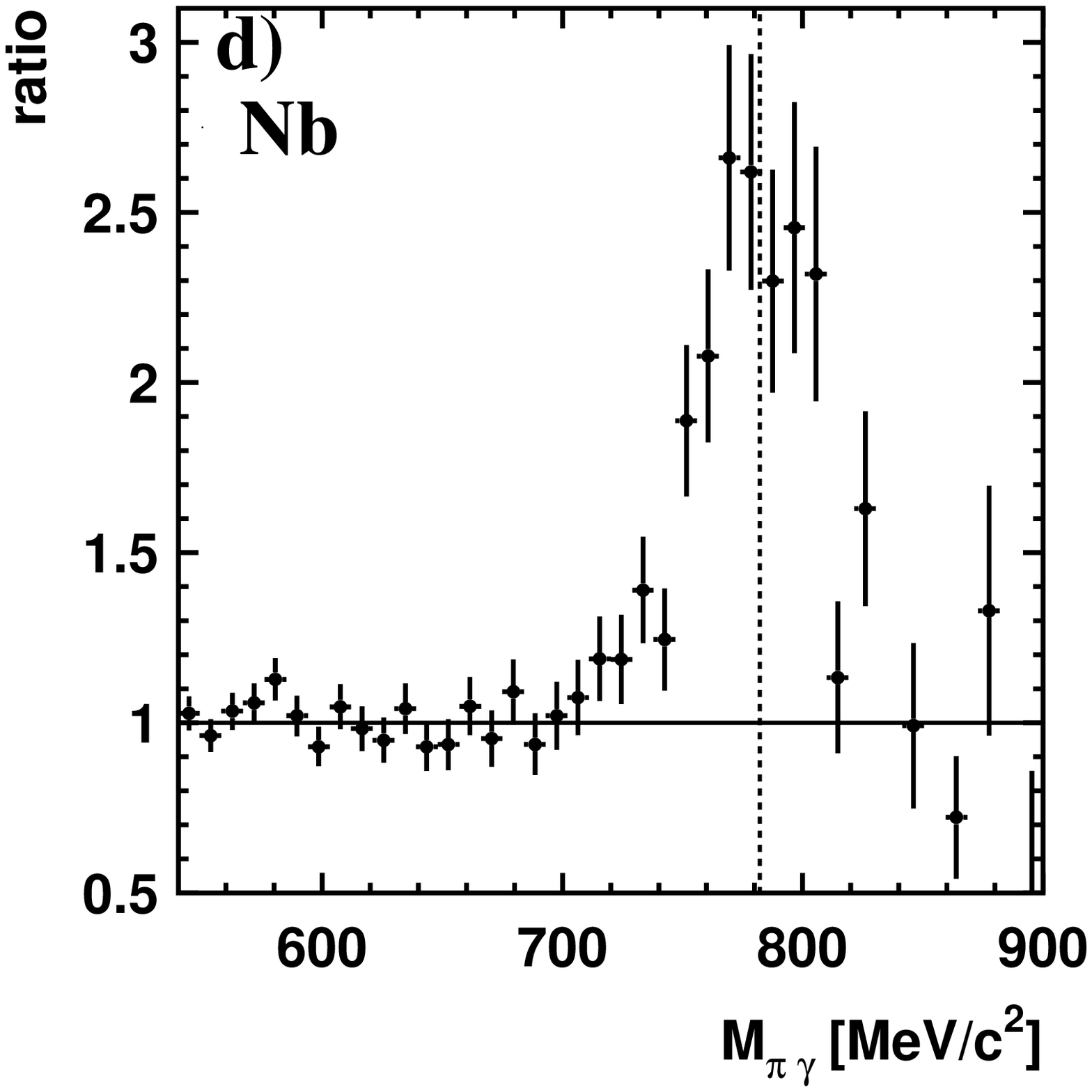}}
\caption{(Color online) a) $\pi^0\gamma $ signal (solid curve) and background spectrum for
   the $C$ target deduced from events with 4 neutral and 1 charged hit. b)
   Ratio of the distributions in a), reflecting the energy dependence of the probability for detecting 3 out of 4 photons relative to detecting all 4 photons per event.  The red (solid) curve is a fit to the ratio ignoring the $\omega$ mass range. The blue (dashed) histogram represents a Monte Carlo simulation. c) The $\pi^{0} \gamma$ signal spectrum and the corrected and  normalized background spectrum for the $Nb$ target. The solid curve
   represents a fit to the background distribution. d) Ratio of the $\pi^{0}
   \gamma$ signal spectrum to the background spectrum for the $Nb$ target generated
   from events with 4 neutral and 1 charged hit.}
\label{fig:C_Nb}
\end{figure*}
Gallmeister et al. \cite{Gallmeister}
argue that a search for medium effects would be much more promising for
incident photon energies below or near the photoproduction threshold on a
free nucleon of $E_{\gamma}^{lab} = 1109$ MeV. New calculations along these lines illustrate in
Fig.\ref{fig:GiBUU} the expected sensitivity of the $\omega$ signal to various
in-medium changes, such as mass shift with and without collisional broadening for two different
energy ranges. It is seen that the lower-energy window indeed leads to a more pronounced -- though not
dramatic -- sensitivity than the higher-energy one. This relatively weak sensitivity is to a large extent
simply a consequence of the density profile of the nucleus that spans all densities from 0 to $\rho_0$ and thus smears any density-dependent signal. Assuming for the sake of the argument a density profile with a constant density of $0.6 \rho_0$ - roughly corresponding
to the average density in nuclei - and a sharp fall off at the surface the dash-dotted line
in Fig.\ref{fig:GiBUU} is obtained; here the in-medium signal is significantly stronger. For a realistic nuclear density profile contributions to the spectral function from the surface dominate, suppressing contributions from higher density regions \cite{Mosel:2010mp}.

In both energy windows a tail towards lower
masses is predicted for the scenario of a dropping $\omega$ mass.
This tail is due to
$\omega$ mesons which are produced off-shell within the nucleus. In \cite{Gallmeister,Muehlich_PhD} an even stronger enhancement in the low mass tail region was obtained.
This calculation used a phenomenological method for the off-shell propagation while the present results are based on the theoretical framework provided by Leupold \cite{Leupold} and Juchem and Cassing \cite{Juchem} who have derived equations of motion for the testparticles that represent the spectral function in the transport simulation starting from the general Kadanoff-Baym equations \cite{KB}. While the Leupold and Juchem-Cassing method has a firm theoretical basis it is nevertheless useful to remember that it relies on the assumption of adiabaticity, i.e. small gradients in space and time of all physical properties (potential, density, spectral function). Non-adiabatic effects such as those at level-crossings are not taken into account.

\section{Experimental set up}
The experiment was performed at the ELSA electron accelerator facility 
\cite{Husmann_Schwille,Hillert} at
the University of Bonn, using the Crystal Barrel/TAPS detector set up which
provides an almost complete coverage of the full solid angle. The
\begin{figure*}
 \resizebox{1.0\textwidth}{!}
  {
   \includegraphics{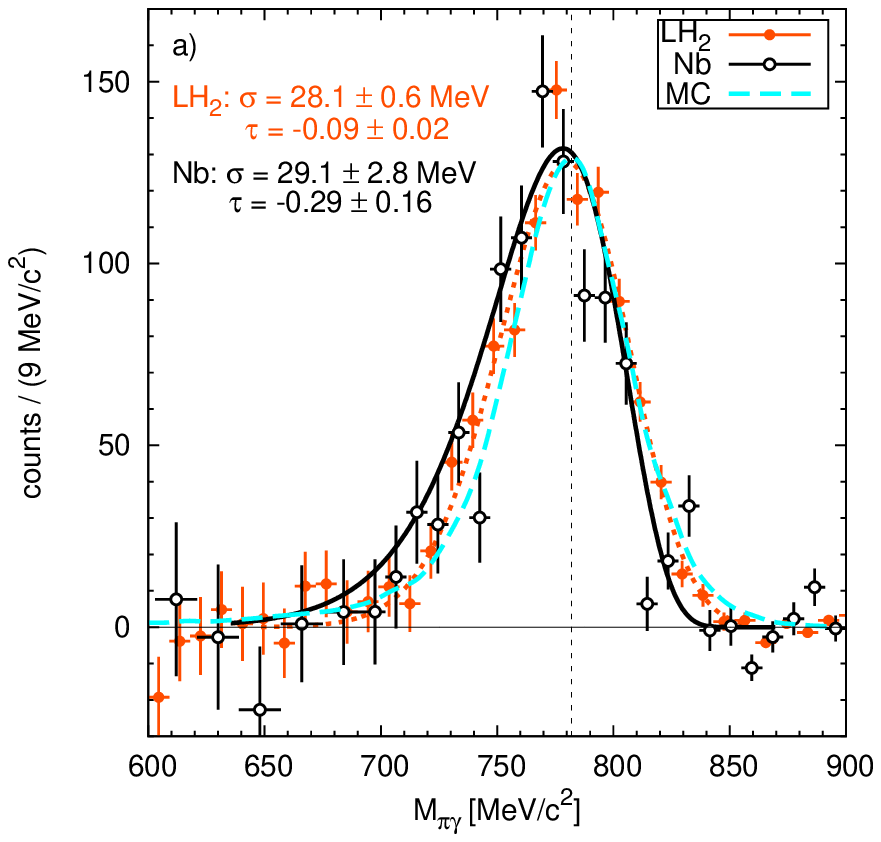}
   \includegraphics{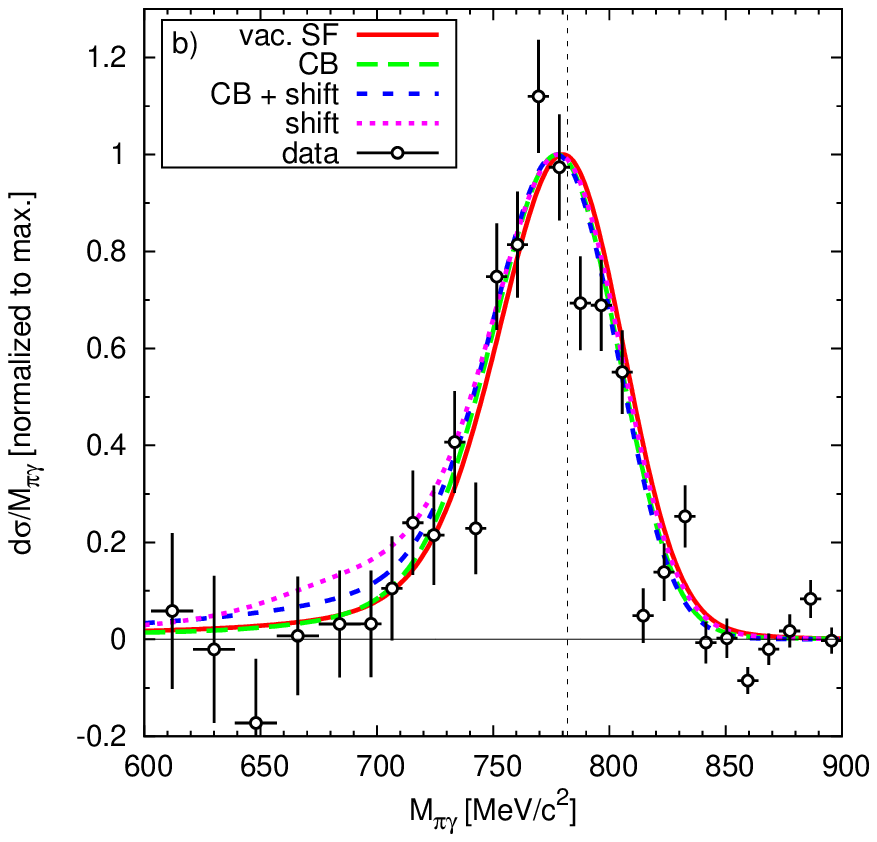}
  }
\caption{(Color online)  a.)  $\omega$ signal (solid points) for the $Nb$ target (1 mm thick) and incident
 photon energies from 900 - 1300 MeV. The errors are purely statistical. Systematic errors introduced by the background subtraction are of the order of 5$\%$ (see Fig.\ref{fig:C_Nb}b.  
 A fit curve to the data points (see text) is shown
 in comparison to the $\omega$ lineshape measured on a 53 mm long $LH_2$ target  and a Monte Carlo simulation;  b.) $\omega$ signal  for the $Nb$ target in comparison to recent GiBUU simulations for the following scenarios: no medium modification (solid), in-medium broadening of $\Gamma_{coll}$ = 140 MeV at nuclear saturation density (long dashed), an additional mass shift by -16$\%$ (short dashed) and mass shift without broadening (dotted). The signals are folded with  with the detector response given by  Eq. \ref{novis_1} with the parameters $\sigma = $ 25 MeV and $\tau$ = - 0.09 and take into account a $1/E_{\gamma}$ weighting of the bremsstrahlung spectrum. }
\label{fig:omega_GiBUU}
\end{figure*}
features of this calorimeter system and its capability for the detection of
multi-photon final states have been described in detail elsewhere
\cite{Aker,Novotny,Gabler}. Tagged photon beams of 900 - 1300 MeV were generated by
bremsstrahlung and impinged in subsequent runs on $LH_2$, $C$ and $Nb$ targets with thicknesses of 53 mm, 20 mm and 1 mm (30 mm diameter), respectively.
The running conditions were the same as described in \cite{Nanova}.

\section{Analysis}
\label{sec:ana}

The analysis of the data follows exactly the scheme described in detail in
\cite{Nanova}. In fact, the data discussed here represent a subset of the data published in 
\cite{Nanova}. $\omega$ mesons were reconstructed in the reaction 
$\gamma A \rightarrow (A-1) p \omega \rightarrow (A-1) p \pi^0 \gamma \rightarrow (A-1) p \gamma \gamma \gamma$ from events with 3 photons and 1 proton in the final state.
Fig.~\ref{fig:C_Nb}a shows the $\pi^0\gamma$ invariant mass
spectrum for the carbon target for events with 1 charged and 3 neutral hits in the
 system. An $\omega$ signal is observed on a steeply falling
background. As discussed in \cite{Nanova} this
background stems dominantly from  $\pi^0\pi^0$ and also $\pi^0\eta
\rightarrow 4\gamma$ final states where due to shower overlap in the
detector or detector inefficiencies 1 out of the 4 photons escapes detection.
Fig.~\ref{fig:C_Nb}a also shows the $\pi^0 \gamma$ invariant mass spectrum for
events with 1 charged and 4 neutral hits where 1
neutral hit has been arbitrarily omitted to simulate the background in the
3 neutral and 1 charged spectrum.
The slopes of the two spectra are different, reflecting the energy dependence of
the probability  to register only 3 out of 4 photons relative to detecting all 4 photons. The ratio of the
two spectra in Fig.~\ref{fig:C_Nb}a is given in Fig.~\ref{fig:C_Nb}b. This ratio is a smooth function of the invariant mass as verified by Monte Carlo simulations, also shown in Fig.~\ref{fig:C_Nb}b.
Since no strong in-medium effects are expected for a light nucleus like carbon this
ratio can be applied to correct the $\pi^0 \gamma$ background spectrum derived from events with 1 charged  and 4 neutral hits measured for the $Nb$ target. Fig.~\ref{fig:C_Nb}c shows the
$\pi^0 \gamma$ invariant mass spectrum for $Nb$ together with this background
distribution after applying this correction. The normalization of the
background spectrum is done by requesting the same number of events in the
mass range from 400 - 900 MeV, excluding the counts in the $\omega$ peak which
account for only 2 $\%$ of the total yield in this mass range. It is important to note that
this determination of the background in magnitude and shape does not pay any attention to the
$\omega$ signal region. The background in the (3 neutral +1 charged) signal
spectrum is well reproduced by the corrected (4 neutral + 1 charged) spectrum.
This is demonstrated by the ratio of the two spectra shown in
Fig.~\ref{fig:C_Nb}d.
In the mass range from 400 - 700 MeV the average deviation from unity is 5 $\%$.
This uncertainty determines the systematic error in the background subtraction.

\section{Comparison to the LH$_2$ reference measurement and to GiBUU simulations}
After subtraction of the background spectrum from the $\pi^0 \gamma$ signal
spectrum the $\omega$ lineshape shown in Fig.~\ref{fig:omega_GiBUU}a is
obtained for the $Nb$ target. The experimental distribution has been fitted using the Novosibirsk function \cite{Novosibirsk}:

\begin{equation}
f(x) = A \cdot \exp\left[{-\frac{1}{2} \left(\frac{\ln{q_{x}}}{\tau}\right)^2 + \tau^{2}}\right] \label{novis_1}
\end{equation}
where
\begin{equation}
q_{x} = 1+ \frac{(x-x_0)}{\sigma} \cdot \frac{\sinh(\tau\sqrt{\ln{4}})}{\sqrt{\ln{4}}} \label{novis_2}
\end{equation}

Here $A$ is the amplitude of the signal, $x_0$ is the peak position, $\sigma$ is
FWHM/2.35 and $\tau$ is the asymmetry parameter. This function takes into account
the tail in the region of lower invariant masses resulting from the energy
response of the calorimeters. The fit  in the mass range 700 to 820 MeV is compared with the $\omega$ signal measured on the $ LH_2$ target and with a Monte Carlo simulation of the $\omega$ signal in Fig.~\ref{fig:omega_GiBUU}a. It should be noted that the 53 mm length of the $LH_{2}$ target leads to an increase in $\omega$-signal width by less than 10 $\%$. The agreement between the $\omega$ signal on the $LH_2$ target with the Monte Carlo simulation demonstrates that the detector response is under control. The fit parameters are $\sigma$ = (29.1 $\pm$ 2.8) MeV, $\tau$ = -0.29 $\pm$ 0.16 for $Nb$ and $\sigma$ = (28.1 $\pm$ 0.6) MeV, $\tau$ = -0.09 $\pm$ 0.02 for $LH_{2}$, respectively. The fit of the data with the function of eqs.(1),(2) yields a $\chi^2$/DoF = 18.95/12 with a $\chi^2$- probability of 9.0$\%$ while a fit of the data with the 
$\omega$ line shape measured on the $LH_2$ target gives a $\chi^2$/DoF = 32.8/15, corresponding to a $\chi^2$-probability of 0.5$\%$.  Nevertheless, in view of the systematic and statistical uncertainties no significant deviation from the reference signals is claimed. Higher statistics will be needed to draw any conclusion. Corresponding data have been taken at MAMI-C using the Crystal Ball/TAPS set up. The analysis is ongoing.

In Fig.~\ref{fig:omega_GiBUU}b the measured $\omega$ signal  is compared to  predictions of
transport calculations using the GiBUU model \cite{GiBUU} for the same scenarios as in Fig.~\ref{fig:GiBUU}. While all the curves
seem to underestimate the data slightly on the low mass side of the $\omega$ peak,
the experimental data obviously do not allow to distinguish between the various theoretical
scenarios, in contrast to initial expectations.

\section{Conclusions}
The photoproduction of $\omega$ mesons on $LH_2$, $C$ and $Nb$ targets
has been measured for incident photon
energies from 900 - 1300 MeV, i.e. near the photoproduction threshold on a free
nucleon of $E_{\gamma}^{lab}$ = 1109 MeV. The experimentally observed
$\omega$ signal does not allow to distinguish between various in-medium scenarios
which - also near threshold - lead all to only a weak tail at low invariant masses. Access to the in-medium spectral function of vector mesons
is thus very limited, mainly due to the dependence of the in-medium properties such as mass and width on the nuclear density
\cite{Mosel:2010mp,Lehr} and the inherent density-smearing caused by the density profile of nuclei. On the other hand,  transparency measurements \cite{Kaskulov,Muehlich} can give at least access to the imaginary part of the in-medium self energy of the hadron. Another promising tool could be the measurement of excitation functions \cite{Om-ex}. Such experiments are presently being analyzed.

\section{Acknowledgements}
We thank the scientific and technical staff at ELSA and the collaborating
institutions for their important contribution to the success of the
experiment. We gratefully acknowledge detailed discussions with
S. Leupold and P. M\"uhlich. This work was supported financially by the {\it
 Deut\-sche Forschungs Gemeinschaft}  through SFB/TR16, by the {\it Schweize\-ri\-scher Nationalfond} and by the {\it Stichting voor Fundamenteel Onderzoek der Materie (FOM)}
and the {\it Nederlandse Organisatie voor Weten\-schappelijk Onder\-zoek (NWO)}.

\end{document}